# Universality of the de Broglie-Einstein velocity equation

Yusuf Z. Umul

*Abstract*—The de Broglie-Einstein velocity equation is derived for a relativistic particle by using the energy and momentum relations in terms of wave and matter properties. It is shown that the velocity equation is independent from the relativistic effects and is valid also for the non-relativistic case. The results of this property is discussed.

## I. INTRODUCTION

In this paper, we will discuss the universality of an important relation named as the de Broglie-Einstein velocity equation [1]. This relation can be defined as

$$v_g v_p = c^2 \quad (1)$$

where $v_g$ and $v_p$ are the group and phase velocities of the de Broglie waves [2]. $c$ represents the speed of light. The general opinion about this equation suggests that it is not valid for the non-relativistic particles. The reason of this proposal is the relation between the phase and group velocities of a quantum particle which can be given by

$$v_g = 2v_p \quad (2)$$

for the non-relativistic case. It is obvious that Eq. (2) contradicts with Eq. (1). In fact this equation is obtained from the non-relativistic energy and momentum relations of

$$E = \frac{1}{2} m_0 v_g^2 \quad (3)$$

and

$$p = m_0 v_g \quad (4)$$

for $E$ and $p$ are the kinetic energy and momentum of the particle. $m_0$ is the rest mass.

Our proposal is the universality of Eq. (1) and the contradiction of Eq. (2) with the special theory of relativity. It is the aim of this paper to give a brief physical and mathematical proof of the suggestion and discuss the results of this approach. We will use the ideas of de Broglie and the special theory of relativity [4, 5] in the construction of our method.

Y. Z. Umul: Author is with the Cankaya University, Yuzuncu Yil, Balgat, Ankara 06530 TURKIYE (corresponding author's phone: +903122844500; e-mail: yziya@ cankaya.edu.tr).

## II. THEORY

One of the most important features of the quantum theory is the duality of wave and particle which was put forward by the works of Planck [6], de Broglie [2, 3] and Einstein [7]. The relativistic equations of energy and momentum can be defined by

$$E = mc^2 \quad (5)$$

and

$$p = mv_g \quad (6)$$

where $m$ is the relativistic mass of the quantum particle. It can be written as

$$m = \frac{m_0}{\sqrt{1 - \dfrac{v_g^2}{c^2}}} \quad (7)$$

according to the special theory of relativity [4]. The relations of energy and momentum can be given by

$$E = \hbar k \quad (8)$$

and

$$p = \hbar w \quad (9)$$

when the wave properties of a quantum particle is considered. $\hbar$ is the angular Planck's constant. $k$ and $w$ are the wave-number and angular frequency, respectively. It is important to note that the dependence to the relativity of the energy and momentum arises from the terms of $m$. We will define a quantity of $\alpha$ by taking the ratio of

$$\alpha = \frac{E}{p}. \quad (10)$$

it is obvious that $\alpha$ is independent from the relativistic effects since it is not a function of the relativistic mass. It can be defined as

$$\alpha = \frac{c^2}{v_g} = \frac{w}{k}. \quad (11)$$

Since the ratio of the angular frequency and wave-number is equal to the phase velocity, Eq. (11) directly leads to Eq. (1). In fact this derivation of Eq. (1) proves that it is independent from the relativistic effects since this equation is obtained by taking the ratio of two quantities which depend on relativity in the same way.

## III. INVESTIGATION OF THE NON-RELATIVISTIC EFFECTS

In this section we will study the non-relativistic equations of the energy and momentum by considering the results, obtained in Section II. Equation (7) can be rewritten as

$$m = \frac{m_0}{\sqrt{1 - \frac{v_g}{v_p}}} \quad (12)$$

when Eq. (1) is taken into account. The non-relativistic mass can be found by

$$m \approx m_0 \left(1 + \frac{v_g}{2v_p}\right) \quad (13)$$

for $v_g \ll v_p$. Equation (2) also contradicts with this condition. According to Eq. (2) which is widely accepted in the literature, it is not possible to get the approximation of Eq. (13) since the condition of $v_g \ll v_p$ is not satisfied. Now we will define the relations of energy and momentum as

$$E \approx m_0 c^2 \left(1 + \frac{v_g}{2v_p}\right) \quad (14)$$

and

$$p \approx m_0 v_g \left(1 + \frac{v_g}{2v_p}\right) \quad (15)$$

for the non-relativistic approximation. It is apparent that the value of $\alpha$ does not change for this case. Equation (3) can be obtained by subtracting the term of $m_0 c^2$ from Eq. (14). But the momentum must be taken as

$$p \approx m_0 \frac{v_g^2}{2v_p} \quad (16)$$

for this case. Otherwise the contradictive result of Eq. (2) will be obtained.

## IV. CONCLUSION

In this paper we showed that the straightforward usage of Eqs. (3) and (4) yields a contradiction which exposes itself with Eq. (2). It is shown that this equation contradicts with the special theory of relativity. It is mentioned and proved that the de Broglie-Einstein velocity equation is universal and is valid for the non-relativistic cases since it does not depend on the relativistic case. It must also be put forward that this equation is obtained from the fundamental equations of the quantum theory and special theory of relativity. For this reason it is reasonable to interrogate the correctness of Eq. (2) instead of eliminating Eq. (1) for the mechanics of the non-relativistic particles. It is important to note that this result also affects the correctness of the differential equations of the quantum theory like the Schrödinger equation.